\documentclass[a4,aps,amsmath,floatfix,nofootinbib,10pt]{revtex4} 
\usepackage{graphicx} \usepackage{color} \usepackage{enumerate} 
\newcommand{\be}{\begin{equation}} \newcommand{\ee}{\end{equation}} 
\newcommand{\ba}{\begin{array}} \newcommand{\ea}{\end{array}} 
\newcommand{\bea}{\begin{eqnarray}} \newcommand{\eea}{\end{eqnarray}} 
\newcommand{\bdm}{\begin{displaymath}} 
\newcommand{\edm}{\end{displaymath}}

\begin{document}

\title{Surprising variants of Cauchy's formula for mean chord length.}

\author{Prabodh Shukla}
\affiliation {Department of Physics, North Eastern 
Hill University, Shillong-793022, India} 
\author{Diana Thongjaomayum} 
\affiliation{ Center for Theoretical Physics of Complex Systems, 
Institute for Basic Science (IBS), Daejeon 34051, Republic of Korea}

\begin{abstract} 

We examine isotropic and anisotropic random walks which begin on the 
surface of linear ($N$), square ($N \times N$), or cubic ($N \times N 
\times N$) lattices and end upon encountering the surface again. The 
mean length of walks is equal to $N$ and the distribution of lengths 
$n$ generally scales as $n^{-1.5}$ for large $n$. Our results are 
interesting in the context of an old formula due to Cauchy that the 
mean length of a chord {\color{blue}{through}} a convex body of volume 
$V$ and surface $S$ is proportional to $V/S$. It has been realized in 
recent years that Cauchy's formula holds surprisingly even if chords 
are replaced by irregular insect paths or trajectories of colliding gas 
molecules. The random walk on a lattice offers a simple and transparent 
understanding of this result in comparison to other formulations based 
on Boltzmann's transport equation in continuum.

\end{abstract}

\maketitle

%%%\section{Introduction}

Augustin-Louis Cauchy (1789-1857)~\cite{cauchy} derived a number of 
mathematically rigorous results with far reaching applications in 
physics. Cauchy's theorem for line integrals of holomorphic functions 
in the complex plane is a prime example. Another result which 
{\color{blue}{has found numerous applications in}} recent years 
concerns the mean length of a chord inside a $d-$dimensional spheroid 
body. Cauchy's formula states that the average chord length (over an 
ensemble of straight lines $AB$ joining a randomly chosen pair of 
points $A$ and $B$ on the inside surface of the body) is proportional 
to the volume $V$ of the body divided by its surface $S$. The constant 
of proportionality $\eta_d$ depends on the dimension; $\eta_3=4$ for a 
sphere and $\eta_2=\pi$ for a circle. The simplicity of this result is 
appealing although not too surprising because the volume and the 
surface are the only free parameters in the problem and $V/S$ has the 
dimension of a length. What is surprising is that the result seems to 
hold even if the straight chord $AB$ is replaced by a random zig-zag 
trajectory of a gas molecule entering $V$ at point $A$ and leaving it 
at point $B$ (first exit). Even more surprising is the apparent 
independence of the result from the details of collisions between 
molecules. The mean chord length plays a key role in several practical 
problems including {\color{blue}{neutron scattering with nuclei}} 
~\cite{dirac}, {\color{blue}{stereology}} ~\cite{underwood}, image 
analysis ~\cite{serra}, and understanding heterogeneous materials 
~\cite{torquato}. As may be expected, a result with random walks 
replacing chords would have a much greater applicability. It is 
observed that Cauchy's formula applies to some biological problems as 
well pertaining to insect behavior. The average distance travelled by 
an ant between its entry into a circular domain and the first exit from 
it is proportional to the radius of the circle ~\cite{blanco}. These 
and other potential applications have inspired several theoretical 
studies in recent years in generalizing the original Cauchy's formula.

Extant studies assume that trajectory of a {\color{blue}{particle}} 
between its first entry and exit from $V$ comprises $n$ line segments 
of lengths $\ell_1,\ell_2,\ldots,\ell_n$ oriented randomly with respect 
to each other. The molecule moves at constant speed in continuum space 
and suffers $n \ge 0$ collisions with other molecules during its stay 
in $V$. Analysis based on Feynman's path integral ~\cite{zoia} as well 
as simplified versions of the Boltzmann transport equation 
~\cite{mazzolo} leads to similar conclusions. It predicts the average 
length of the trajectory $<L> = <\ell_1+\ell_2+\ldots+\ell_n> = \eta_d 
R$, where $R$ is the radius of the bounding $d-$dimensional sphere for 
isotropic random walks. If the average length of a segment $\ell_i$ 
between two successive collisions $\lambda = <\ell_i> = <L> / n$ exists 
in the limit $n \to \infty$, a mean-field like solution predicts that 
segment lengths $\ell_i$ are distributed exponentially according to the 
probability $P(\ell_i) = exp(-\ell_i/\lambda) / \lambda$. The result 
$<L>=\eta_d R$ also holds for an arbitrary distribution of $\ell_i$ if 
a constraint is imposed between the distribution of the first step 
$\ell_1$ that injects the walker inside $V$ and the distribution of 
subsequent step lengths ~\cite{mazzolo}. In the present paper, we study 
the problem on an $N \times N$ square lattice. Here each step of the 
walk is of unit length and the total length of the walk is simply the 
number of steps $n$ between entry into the lattice and first exit from 
it. Our main finding is that the key feature of Cauchy's formula holds 
for isotropic as well as anisotropic random walks on the lattice. The 
average length of the walk $<n>$ scales linearly with $N$ but 
surprisingly the distribution of walk lengths $n$ in different 
realizations of the walk follows a power law. In the following we 
present numerical results as well as theoretical support for them. The 
results may be easily generalized for a $d-$dimensional hypercubic 
lattice for $d>2$. The results extend Cauchy's formula on lattices but 
more interestingly provide a simple intuitive understanding of the same 
based on a one-dimensional random walk with two absorbers.

%%%\section{Simulations}

Fig.1 illustrates a computer generated isotropic random walk on a small 
$10 \times 10$ lattice. In this particular realization the walker takes 
a total of $n=68$ steps through 27 lattice points. Starting at the 
entry point $(6,1)$ on the edge of the lattice, she takes the first 
step to $(6,2)$ and randomly walks on for a total of 68 steps till she 
reaches the boundary again at $(1,6)$ and terminates the walk. The 
first and last steps are necessarily perpendicular to the boundary but 
other steps occur with equal probability in any of the four directions. 
The directions of the first eight and the last five of the sixty-eight 
steps are shown by arrows. The remaining steps are left unmarked 
because these are traversed back and forth several times making 
overlapping loops of various lengths. This is not an artifact of the 
small system size but rather a general feature of restricted (surface 
to surface) random walks. The walks tend to be loopy and localised in 
crowded neighborhoods which are separated from each other by longer and 
less loopy paths.{\color{blue}{ On account of the number of loops that 
a walker may make in a localized region, the size of the localized 
region is not a true indicator of the length of walk through it. This 
feature endows the system with a scale invariant property.}} It gives 
rise to a power-law distribution for key quantities. In our simulations 
on hypercubic lattices of $N^d$ sites we focus on two quantities: (i) 
$<n>$, the mean length of the walk, and (ii) $P(n;\{p\};N)$, the 
distribution of lengths. Here $\{p\}$ is a set of $2d$ probabilities 
for moving in different directions on the lattice. For example, an 
isotropic walk on a square lattice is denoted by 
$p=0.25;0.25;0.25;0.25$. It serves to distinguish between symmetric and 
asymmetric walks. Our main findings are (i) $<n>$ scales linearly with 
$N$ for symmetric and asymmetric walks, and (ii) $P(n;\{p\};N)$ shows 
power-laws for symmetric and a few asymmetric cases as well. The result 
for $<n>$ is in conformity with the general idea of Cauchy's formula. 
The power-law distribution is in contrast to extant studies in 
continuum where a mean field solution produces an exponential 
distribution.

{\color{blue}{ Fig.1 depicts just a single walk for illustrative 
purposes. In simulations on larger systems presented below, each point 
on the boundary (except corners on square and cubic lattices) was 
assigned an equal probability of being chosen as the starting point of 
the walk. The direction of the first step was restricted to the nearest 
neighbor of the starting point lying inside the boundary. The corner 
sites were excluded because they did not have a nearest neighbor inside 
the boundary. No restriction was placed on the direction of the walk 
after the first step. Results were obtained by (a) averaging over a 
large number of randomly selected starting points on the boundary, and 
(b) averaging over all points on the boundary as starting points. As 
may be expected, results in the two cases are indistinguishable from 
each other on the scale of the figures. The numerical procedure 
outlined above is a natural realization on lattices of the two 
conditions for the validity of Cauchy's formula ~\cite{mazzolo}: (i) 
the starting position should be uniformly distributed over the 
boundary, and (ii) the starting direction should satisfy isotropic 
incident flux condition.}}

Fig.2 shows the result for $<n>$ on an $N^d$ lattice for $10 \le N \le 
10^3$, $d \le 3$, and different cases of symmetric as well as 
asymmetric walks. We find $<n>$ to be proportional to $N$ in each case 
within numerical errors. The constant of proportionality is unity for 
symmetric walks but different from unity for asymmetric walks for 
reasons that are simple to understand and explained in the following. 
After scaling by appropriate weight factors, the data for all cases 
collapse on a single line as shown in Fig.2. For a symmetric walk, the 
walker moves to any of its nearest neighbors with equal probability. In 
an asymmetric walk, the probabilities to go to different neighbors are 
different. On a square lattice, we consider two cases of asymmetry for 
exit from site $(i,j)$. Case-I: the probability to go to $(i-1,j)$ or 
$(i+1,j)$ is equal to 0.10 but the probability to go to $(i,j-1)$ or 
$(i,j+1)$ is equal to 0.40. Case-II: the probability to go to $(i-1,j)$ 
is equal to 0.10, and to $(i+1,j), (i,j-1)$, or $(i,j+1)$ is equal to 
0.30 in each direction. In other words, the asymmetry in Case-I is 
between left-right ($x$-axis) and up-down ($y$-axis) steps. In Case-II 
there is a asymmetry along the $x$-axis in addition to asymmetry 
between the two axes. The probabilities to go to different neighbors 
add up to unity ensuring the walker does move to a new site in each 
step. The proportionality of $<n>$ to $N$ and the constants of 
proportionality can be understood by focusing on the one dimensional 
case. We shall return to it shortly after presenting the results for 
the distribution of walk lengths.

The distributions $P(n;\{p\}; N)$ for fixed $\{p\}$ and different $N$ 
are qualitatively similar. The range of $n$ increases with increasing 
$N$ but the scatter in the data reduces when larger number of 
realizations of the walk are used to calculate the distribution.  We 
also note that $P(n;\{p\}; N)$ remains finite for $n >> N$. Thus even 
for $N \le 10^3$, simulations generate huge data files, particularly so 
for symmetric walks. Fig.3 is drawn using reduced data based on 
logarithmic binning of the raw data into a reasonable number of bins 
that preserve the trends of the raw data. The raw data for Fig.3 was 
obtained for $N=10^3$ and more than $10^6$ realizations of the walk. 
The important features of Fig.3 are the following. For symmetric walks, 
$P(n;\{p\};N)$ shows a power-law decrease with increasing $n$ over 
several decades. Within numerical errors, $P(n;\{p\};N) \approx 
n^{-1.5}$ for large $n$. For an asymmetric walk belonging to Case-I, 
the distribution of walk lengths is qualitatively similar to the one 
for the symmetric walk. However it is different for the asymmetric 
Case-II. In this case we do not see a clear power-law although the 
departure from the power-law does not look very pronounced on the scale 
of Fig.3. We can understand Case-II more clearly by referring to Fig.4 
which shows the results for symmetric and asymmetric walks on a $1d$ 
lattice of length $N$.

%%%\section{Theory}

In one dimension, the problem can be solved analytically 
~\cite{ellis,feller}. On a linear lattice $1, 2,\ldots, m, \ldots N-1, 
N$ with $p$ the probability of moving towards $N$, $q=1-p$ the 
probability of moving towards $1$, sites $1$ and $N$ being absorbers, 
the average number of steps $S_m$ required to start from $m$ and get 
absorbed at 1 is given by,

\be S_m = \frac{m}{q-p}-\frac{N}{q-p} .\left[ \frac{1-r^m}{1-r^N} 
.\right] \mbox{( if $p > q; r=q/p$ )} ; S_m=m (N-m) \mbox{ (if $p = q = 
1/2$)} \ee The probability that it takes exactly $n$ steps to get 
absorbed into attractor-1 is given by,

\be U(1,n) =\frac{1}{N} 2^n p^{\frac{n-1}{2}} q^{\frac{n+1}{2}} 
\sum_{\nu=1}^{N-1} \cos^{n-1}{\frac{\nu \pi}{N}}\sin{\frac{\nu 
\pi}{N}}\sin{\frac{\nu \pi m}{N}} \ee

The key points are that the average length of the walk scales linearly 
with $N$, and the probability that a symmetric walk is completed in $n$ 
steps scales as $n^{-1.5}$ for $n\to \infty$. Simulations show that 
these features hold in two and three dimensions as well. The reason is 
as follows. Consider isotropic walks on a square lattice. A walker may 
start from a randomly selected site on an edge, say site $(1,m)$ on the 
top edge (first row). The first step brings her straight down to the 
second row at point $(2,m)$. Thereafter she can move in any direction 
with equal probability taking one step at a time till she hits one of 
the edges. We may imagine that each step is decided by two tosses of a 
coin, the first toss deciding if she would move along a column or a 
row, and the second deciding the direction. Now consider a modified 
walk where she decides to ignore the first toss and stays on the column 
$m$. She uses the second toss to move up or down on column $m$ with 
equal probability. This modified walk is a one-dimensional walk with 
rescaled time. The average number of steps needed to hit either top or 
bottom edge is equal to $N-1$, i.e. the average length of the walk 
scales linearly with $N$. We can also imagine a walk confined to the 
first row moving left or right with equal probability but skipping 
steps in the vertical direction. The average length of the walk before 
it hits the left or the right edge of the square is equal to $m(N-m)$, 
again scaling linearly with $N$. On a $d$-dimensional hypercubic 
lattice we may consider components of the walk along each dimension 
independently and so the qualitative features of the composite walk are 
the same as for each component. This explains why the average length of 
the walk scales linearly with $N$, and the distribution of lengths as 
$n^{-1.5}$ irrespective of the dimensionality of the lattice.

Similar reasoning may be used to understand asymmetric walks. Fig.4 
shows the distribution of lengths for three cases, (i) $1d$ 
asymmetric; (ii) $1d$ symmetric (for comparison); and (iii) $2d$ 
asymmetric. The idea behind Fig.4 is to use the $1d$ example to 
understand the scaling of data for asymmetric walks in Fig.2 which 
make it collapse on a common curve for symmetric walks, and also to 
understand the shape of one $2d$ curve in Fig.3 which does not follow a 
clear power-law. At the risk of some repetition, the data for the $1d$ 
curves in Fig.4 is obtained as follows. On the lattice 
$1,2,\ldots,N-1,N$, start from site-$2$ or site-$(N-1)$ with equal 
probability. Take a step towards $N$ with probability $p$ or towards 
site-$1$ with probability $q=1-p$. Count the number of steps $n$ to 
reach site-$1$ or site-$N$ which ever occurs first. Average the 
distribution of $n$ over $10^7$ realizations of the walk. For $p=0.50$ 
both absorbers are reached equally frequently and the probability of 
reaching any one of them scales as $n^{-1.5}$ for large $n$. In higher 
dimensions as well, the power-law distribution is obtained whenever the 
two absorbers on opposite sides of a coordinate axis have equal 
probability of encounter.

For the $1d$ asymmetric case, the two absorbers are not reached with 
equal probability. Fig.4 shows that the probabilities of hitting 
absorber-$1$ or absorber-$N$ at the end of $n$ steps are in the ratio 
0.25:0.75. The probability of hitting $N$ is larger because the walk is 
biased towards it. To calculate the average length $<n>$ we have to use 
different weight factors for the two absorbers. This is relatively easy 
in $1d$ because there are only two absorbers but gets tedious in higher 
dimensions. A simplification is provided by the fact that the weight 
factors of different absorbers do not depend on the length $n$. Thus we 
may not count the frequency of different absorbers separately but 
simply rescale the cumulative number of instances where the walk hits 
an absorber after $n$ steps. In the two cases of asymmetric walk 
$\{d=2; p= 0.10;0.10;0.40;0.40\}$ and $\{d=2; p= 
0.10;0.30;0.30;0.30\}$, the cumulative frequencies need to be weighted 
by $0.89$ and $1.20$ respectively to make $<n>$ vs. $N$ data collapse 
on a single line as shown in Fig.2.

The probability to hit the absorber $N$ comprises two disjointed parts; 
a part that starts at 0.375 at $n=1$ and decreases rapidly for higher 
odd values of $n$, and another part with a peak around $n \approx 
2000$. The exponentially decreasing part corresponds to a walk starting 
at $N-1$ and getting absorbed at $N$ after $n=1,2,3\ldots$ steps. The 
simulation data agrees with the analytic solution mentioned above but 
the peak is a finite size effect. There are two parameters of interest 
$n$ and $N$. The analytic result applies to $n \le N$. However in 
numerical simulations we may have $n >> N$, and a substantial number of 
walks terminate at $N$ for $n \approx 2N$. This accounts for walks that 
start from site-$2$ and continue towards site-$N$; for $p=.75$, 
$q=.25$, it takes nearly $2N$ steps to cover the distance to $N$. As 
the asymmetry decreases, the following effects set in: (i) rapidly 
decreasing branch tends to a slower power-law decrease and extends to 
higher $n$, (ii) the peak diminishes as well and may not remain 
disjoint from the other part of the curve.

A combination of above effects explains the $2d$ asymmetric case $\{p= 
0.10;0.30;0.30;0.30\}$ in Fig.4 if we consider the following points; 
(i) asymmetry is along the $x$-axis, (ii) there is a peak at $n \approx 
5000$, (iii) $10\%$ of $5000$ steps are taken towards column $1$ and 
$30\%$ towards column $N$, (iv) $5000 \times (0.30-0.10) = 1000 =N$, 
(v) thus a walk starting next to first column reaches the last column 
after $5000$ steps and is terminated, (vi) $P(n;\{p\};N)$ is rather 
close to a power-law because the probability to move along three 
directions is each equal to $0.30$ so there is a good deal of isotropy 
in the system.

%%%\section{Concluding Remarks}

{\color{blue}{
In conclusion, we have studied restricted random walks on bounded 
$d$-dimensional hypercubic lattices of size $N^d$. The walks start and 
end on the surface of the lattice. The starting points are uniformly 
distributed over the surface, and the first step of each walk is 
perpendicular to the surface and directed inwards. These conditions are 
the lattice analogue of uniform and isotropic incident flux associated 
with the Cauchy formula for random walks in continuum space bounded by 
an ellipsoid surface~\cite{mazzolo}. As in the continuum case, the 
average length of the walk is proportional to the volume divided by the 
surface of the lattice. The constant of proportionality depends on the 
geometry of the problem. On the hypercubic lattice the average length 
is equal to $N$ i.e. the closest distance between the face of the 
hypercube where the walk starts and the face opposite to it. Of course, 
not all walks end on the opposite face and even those that may end on 
it have a distribution of lengths. We find the lengths of the walk have 
a power-law distribution with the exponent $-3/2$ within numerical 
errors. The geometry of the hypercube and the condition of uniform and 
isotropic incident flux conspire to pair a shorter walk with a longer 
walk nearly in a detailed balance manner. This leads to an average 
value of the length equal to $N$ as mentioned above. The lengths have a 
power-law distribution if the random walk is symmetric i.e. if it has 
an equal probability of taking a step towards anyone of the $2d$ faces 
of the cube.  The power-law distribution also holds in cases when the 
walk is symmetric between each pair of opposite faces, but not 
necessarily symmetric between different pairs. The power-law is lost 
if the symmetry between opposite faces is lost.

The results presented here may have a broader significance on two 
accounts. Firstly, random walks are used to model a large number of 
statistical physics problems. They are primary models of diffusion 
which is a basic mechanism for the evolution of statistical systems. 
Diffusion is used not only to understand a system's evolution from a 
nonequilibrium to an equilibrium state but also to understand 
equilibrium and steady states. Thus different variants of random walks 
may be useful in understanding the richness of diffusion phenomena 
including anomalous diffusion ~\cite{oliveira}. Recent studies of 
diffusion with stochastic resetting ~\cite{evans} are also interesting 
in this context. Stochastic resetting effectively reduces the domain of 
an otherwise perennial random walk and has a marked effect on its 
likelihood to hit a specified trap and get terminated. The random walks 
studied here are confined to a bounded space and this too has a strong 
effect on their properties. It may be interesting to explore if there 
are any general principles governing the effect of a bounded domain on 
the properties of random walks inside it. However, this is somewhat 
outside the scope of the present study.

Secondly, and not entirely unrelated to the point made above, our study 
highlights the effect of the surface on the dynamics of the system 
bounded by it.}} Often statistical models in the thermodynamic limit 
ignore surface effects and focus on the deep interior of the volume. 
This also includes several boundary value problems outside the realm of 
statistical physics. The role of surface is merely to fix the boundary 
conditions for the differential equations to be solved. The possibility 
of some physical quantities depending only on the ratio of 
{\color{blue}{volume to surface}} but independent of the dynamics 
inside {\color{blue}{volume}} may offer new insights and therefore 
needs further exploration. Even in cases where surfaces play a more 
direct role in the theoretical understanding, Cauchy's formula may 
provide a different viewpoint. For example, some reflection shows that 
it can be used to obtain an alternate understanding of gas laws without 
the stringent assumptions of an ideal gas. We hope this note is a small 
step in this direction.

{\color{blue}{ \acknowledgments{ One of us (D.T.) acknowledges the 
support from Institute for Basic Science in Korea (IBS-R024-D1).}}}

\begin{figure}[ht] 
\includegraphics[width=0.75\textwidth,angle=0]{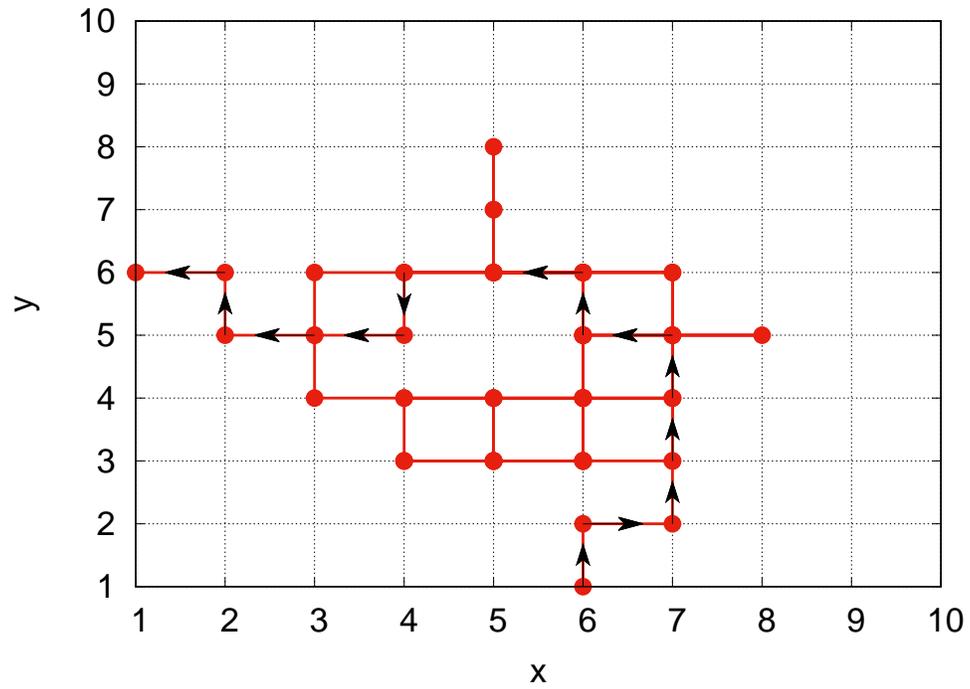} \caption{An 
isotropic random walk on a $10 \times 10$ lattice that starts and ends 
on the boundary. } \label{fig1} \end{figure}

\begin{figure}[ht] 
\includegraphics[width=0.75\textwidth,angle=0]{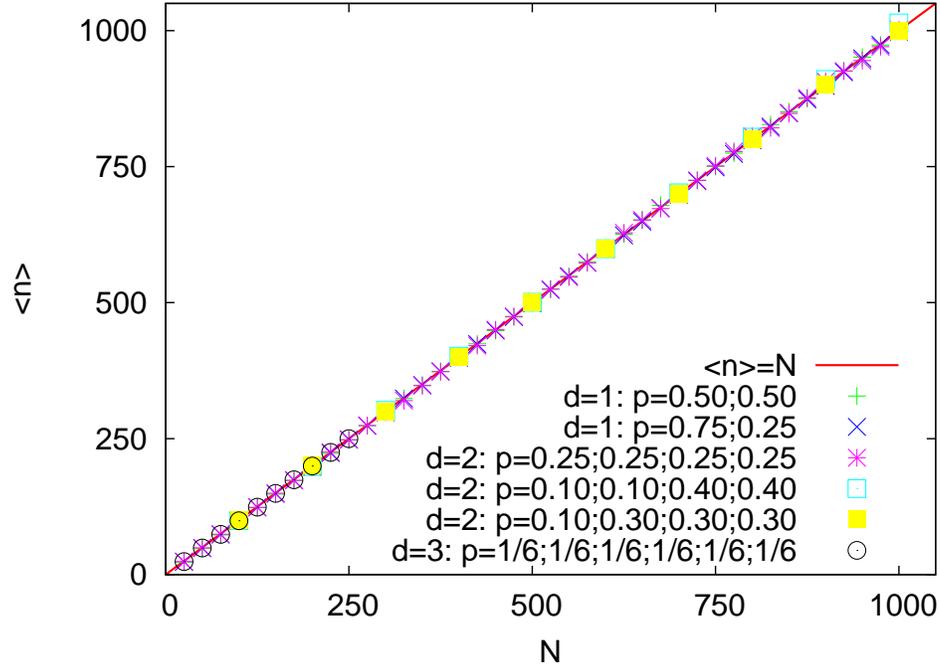} 
\caption{Average length $<n>$ of a random walk starting and ending on a 
surface of a $d$-dimensional hypercubic lattice of linear size $N$. The 
values of $p$ are the probabilities of walking in $2d$ different 
directions. Thus the figure shows three isotropic and three anisotropic 
walks. The data for two anisotropic walks is rescaled (see text) so 
that all cases collapse on a line $<n>=N$.
} 
\label{fig2} 
\end{figure}

\begin{figure}[ht] 
\includegraphics[width=0.75\textwidth,angle=0]{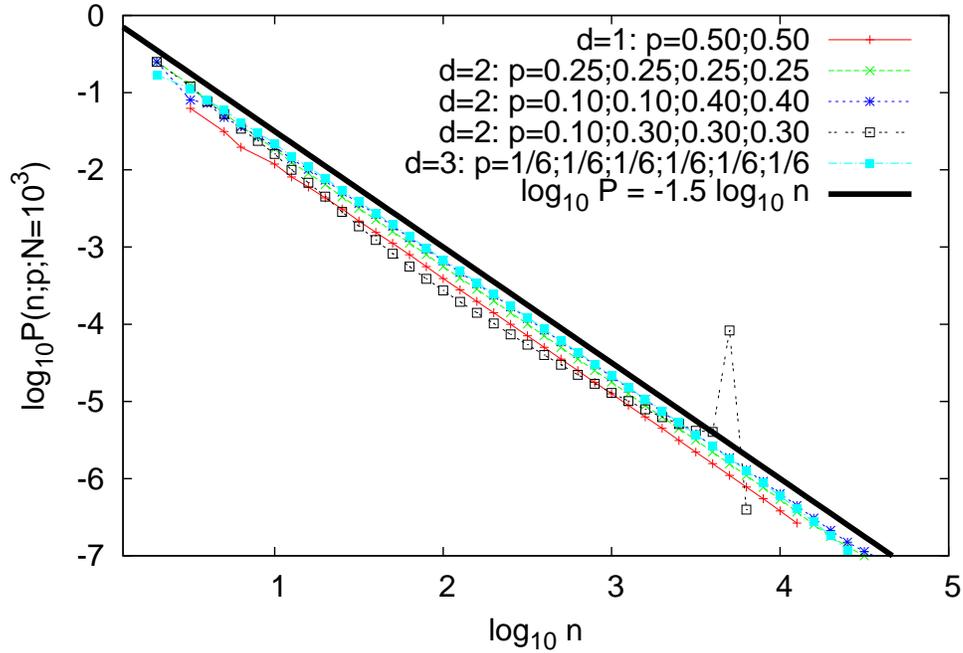} \caption{ 
Probability of a walk comprising $n$ steps shows a power-law behavior 
except for one case shown in open black squares (see text).
} 
\label{fig3} 
\end{figure}

\begin{figure}[ht] 
\includegraphics[width=0.75\textwidth,angle=0]{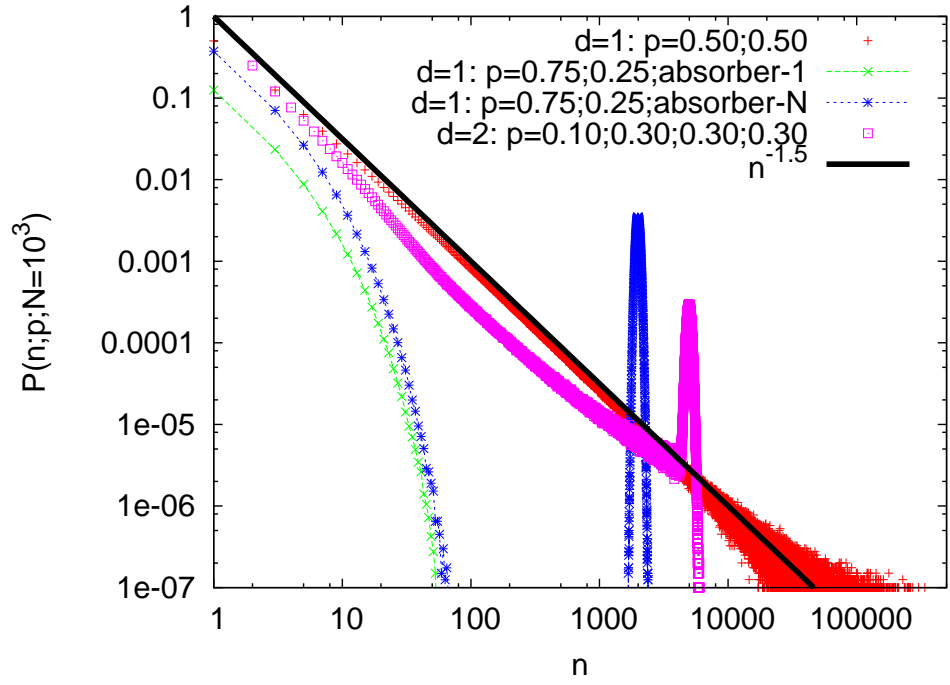} \caption{ 
Similar plot as in Fig.3 but without logarithmic binning of data and 
for selected cases. In $1d$ asymmetric case the probabilities of 
reaching the two absorbers are plotted separately. It explains the 
shape of the anisotropic walk in $2d$ (see text).} \label{fig4} 
\end{figure}

\end{document}